\documentclass[11pt,fleqn]{article}

\usepackage{amsmath} 
\usepackage{amsfonts}
\usepackage{graphicx}
\usepackage{dcolumn}
\usepackage{upgreek}
\usepackage{latexsym,slashed}
\usepackage{amssymb,amsfonts}
\usepackage{cancel}
\usepackage{bm} %% bold mathematics
\usepackage{mathrsfs} %% non gradito a latex2html!
\usepackage{braket}

\def\ds{de Sitter}

\newcommand{\address}[1]{\begin{center}\large #1\end{center}}

\renewcommand\Im{\operatorname{Im}}

\def\beq{\begin{eqnarray}}
\def\eeq{\end{eqnarray}}

\def\R{{\hbox{{\rm I}\kern-.2em\hbox{\rm R}}}} %% real numbers
\def\H{{\hbox{{\rm I}\kern-.2em\hbox{\rm H}}}} %% Hilbert space
\def\N{{\hbox{{\rm I}\kern-.2em\hbox{\rm N}}}} %% natural numbers
\def\C{{\ \hbox{{\rm I}\kern-.6em\hbox{\bf C}}}} %% complex numbers
\def\Z{{\hbox{{\rm Z}\kern-.4em\hbox{\rm Z}}}} %% integers numbers
\catcode`\@=11
\@addtoreset{equation}{section}

\begin{document}
\tolerance=5000

\title{\bf{Vacuum fluctuation of conformally coupled scalar field in FLRW space-times}} 

\author{
Giovanni Acquaviva$\,^{(a)}$\footnote{acquavivag@unizulu.ac.za},
Luca Bonetti$\,^{(b)}$\footnote{bonetti659@gmail.com},
Luciano~Vanzo$\,^{(c)}$\footnote{vanzo@science.unitn.it} and
Sergio~Zerbini$\,^{(c)}$\footnote{zerbini@science.unitn.it}}
\date{}
\maketitle
\address{$^{(a)}$ Department of Mathematical Sciences, \\
 University of Zululand, South Africa}
\address{$^{(b)}$ Universite' d'Orleans, Observatoire des Sciences de l'Univers en Region Centre, France} 
\address{$^{(c)}$ Dipartimento di Fisica, Universit\`a di Trento \\
and TIFPA, INFN Center,  Trento, Italy}
\medskip 
\medskip

\begin{abstract}

The regularized vacuum fluctuation related to a conformally coupled massless scalar field defined on a space-time with dynamical horizon is computed with respect a radially moving observer in a generic flat FLRW space-time. Two simple measurement prescriptions are given in order to remove the ambiguity associated with the short distance singularity of the correlation function. In some cases, it turns out that one is dealing with a ``quantum thermometer", recovering a proposal due to Buchholz \textit{et al.} in order to determine local temperature in the framework of quantum field theory.  In general, by arranging the detector so that it does not register for inertial motion in flat space, the regularized quantum fluctuation may be used as a probe of space-time geometry and, in particular, may provide informations on the Hubble parameter.  As an aside, it is not possible in general to fully decouple the effect of the detector's motion from the universe expansion, a fact that could be interpreted as a kind of Machian effect which can be traced back to the worldwide nature of the vacuum.
\end{abstract}

\section{Introduction}\label{intro}

Relativistic theories of gravity  on flat Friedmann-Lema\^{i}tre-Robertson-Walker (FLRW) space-times have become important in modern cosmology after the discovery of the current cosmic acceleration, the onset of the dark energy issue and after the confirmation of inflationary models. Among the several descriptions of the current accelerated expansion of the universe, the simplest one is the introduction of a small positive cosmological constant in the framework of General Relativity, so that one is dealing with a perfect fluid whose equation of state parameter $\omega=-1$. This fluid model is able to describe the current cosmic acceleration, but also other forms of fluid (phantom, quintessence, inhomogeneous fluids...) satisfying suitable equation of state are not excluded, since the observed small value of cosmological constant leads to several conceptual problems, such as the role of vacuum energy and the coincidence problem.
For this reason, several different approaches to the dark energy issue have been proposed. 
Among them, the modified theories of gravity (see, for example~\cite{Review1}--\cite{rev} and references therein) represent an interesting extension of the Einstein's theory. Unfortunately, a large class of these modified models admit future singularities, the worst being the so called Big Rip singularity \cite{cald,br}.

With regard to quantum fields in curved space-time -- the other \textit{leitmotiv} of this paper -- one of its most important predictions is the Hawking radiation \cite{Haw}.  Several derivations of this effect can be found in literature \cite{dewitt,BD,wald,fulling,igor} and recently the search for ``experimental'' verification making use of analogue models has been pursued by many investigators (see for example \cite{Unruh:2008zz,Barcelo:2005fc}).

In a seminal paper, Parikh and Wilczek \cite{PW}(see also \cite{visser}) introduced a further approach, the so-called tunneling method, for investigating Hawking radiation. A variant of  their method has been also introduced and called Hamilton-Jacobi tunneling method \cite{Angh,tanu,mann}. This method is covariant and enjoys the peculiar feature to admit a generalization to the dynamical case \cite{bob07,sean09,bob09}. For a  recent review, see \cite{Vanzo11} and the references therein, and for a rigorous quantum theoretical approach, see also \cite{moretti}.

Recall that in the tunneling approach, the semiclassical emission rate reads 
\beq\label{Gam}
\Gamma \propto |\mbox{Amplitude}|^2 \propto 
e^{-2\frac{\Im I}{\hbar}} \,. 
\eeq
with $\Im I$ standing for the imaginary part of the classical action.
The leading term in the WKB approximation of the tunneling probability reads
\beq
\Gamma \propto e^{- \frac{2\pi }{\kappa_H} \omega}\,, 
\label{g}
\eeq
 in which an energy $\omega$ of the particle and the surface gravity evaluated at horizon $\kappa_H$ appear. 
From this asymptotic, one obtains the Hawking temperature by the identification $T_H=\frac{\kappa_H}{2\pi}$. The method is quite general and works for a generic stationary black hole.  With appropriate mathematical notions of horizons and surface gravities, the above formula is still valid in the spherically symmetric dynamical case, as shown in \cite{bob07}-\cite{bob09}, but the interpretation of $\kappa_H/2\pi$ as an effective temperature parameter is more delicate and depends on what a local detector on a given trajectory can actually detect. If the horizon emits black body radiation in the surrounding vacuum then by Liouville theorem the distribution function of the radiation is constant along phase space trajectories, and therefore must have a temperature equal to that of the emitting horizon, but red-shifted as predicted by general relativity. If it does not emit as a black body one can still define a local temperature by comparing the radiation density at each point in phase space with the equilibrium Planck density, but such a local temperature will generally depend on frequency and direction. \\
With regard to this ``temperature'' issue, we are particularly interested in the cosmological scenario and we would  like to continue the investigation by making use of quantum field theory to evaluate the fluctuation of the simplest quantum probe at our disposal, namely a conformally coupled scalar field defined on a spherically symmetric space-time with horizons (see \cite{obadia,noi11} and references therein). 

We mainly restrict our analysis to  a flat FLRW space-time, which is a spherically symmetric dynamical space-time admitting in principle several past and future oriented dynamical horizons.  We can now briefly review the general formalism \cite{kodama,hay1,hay2} and the relevant quantities that will be used. 

Any spherically symmetric metric can locally be expressed in the form
\beq
\label{metric}
ds^2 =\gamma_{ij}(x^i)dx^idx^j+ R^2(x^i) d\Omega^2\,,\qquad i,j \in \{0,1\}\;,
\eeq
where the two-dimensional metric
\beq d\gamma^2=\gamma_{ij}(x^i)dx^idx^j
\label{nm}
\eeq
is referred to as the normal metric, $\{x^i\}$ are associated coordinates and $R(x^i)$ is the areal radius, considered as a scalar field in the two-dimensional normal space. A relevant scalar quantity in the reduced normal space is 
\beq
\chi(x)=\gamma^{ij}(x)\partial_i R(x)\partial_j R(x)\,, \label{sh} 
\eeq 
since the dynamical trapping horizon, if it exists, is defined by
\beq 
\chi(x)\Big\vert_H = 0\,, \label{ho} 
\eeq
provided the condition $\partial_i\chi\vert_H \neq 0$ is satisfied.
 One significant scalar in the normal space is given by the interesting proposal due to Hayward \cite{hay1}
\beq
\kappa_H=\frac{1}{2}\Box_{\gamma} R \Big\vert_H\,.
\label{H}
\eeq 
which is a generalization of the usual Killing surface gravity. This is the quantity which appears in the tunneling rate \eqref{Gam}. 
But there is another one, still given by Hayward \cite{hay2}, which is defined by computing on the horizon the quantity
\beq\label{SHk}
{\cal K}_{H}=\frac{1}{2}\sqrt{-n^{\mu}\nabla_{\mu}\theta}\Big\vert_H
\eeq
where $\theta$ is the expansion of an appropriately oriented null geodesic congruence with tangent vector $l^{\mu}$, and $n^{\mu}$ is another future pointing null congruence such that $n\cdot l=-1$. 

As an example, let us consider the  flat FLRW space-time,  the metric is usually   written in the form 
\beq
ds^2=-dt^2+a^2(t)\left(dr^2+r^2\,d\Omega^2 \right)\,,
\eeq
the coordinates are $x=(t, r)$, the areal radius is $R=a(t)\,r$ and the normal metric simply reads 
\beq
d\gamma^2=-dt^2+a^2(t)dr^2\,.
\eeq
Thus, 
\beq
\chi=-(\partial_tR)^2+\frac{1}{a^2(t)}(\partial_rR)^2=0\,,
\eeq
namely the trapping horizon is located at $r_H=\frac{1}{\dot{a}}$ and in term of the areal  radius reads
\beq
R_H = a(t)\,r_H=\frac{1}{H(t)}\,.
\eeq
where the Hubble parameter is given by $H(t)=\frac{1}{a}\frac{d a}{dt}$. The quantity $R_H$ is known as Hubble Sphere, but we may also refer to it as Hubble dynamical horizon in the Hayward's terminology. Thus, it turns out that  $\kappa_{H}=R_H(t)\, {\cal K}_{H}^{2}/2$, so that both definitions contain the same information.

Furthermore, in a spherical symmetric dynamical case, it also is possible
to introduce the Kodama vector field $K$. Given the metric Eq.~(\ref{metric}), the Kodama vector components are
\beq 
 K^i(x)=\frac{1}{ \sqrt{-\gamma}}\,\varepsilon^{ij}\partial_j R\,,
\qquad  K^\theta=0= K^\varphi \label{ko} \;. 
\eeq 
We may introduce the Kodama trajectories, and related Kodama observer, by means of integral lines of the Kodama vector
\beq
\frac{d\, x^i}{d \lambda}= K^i= \frac{1}{ \sqrt{-\gamma}}\,\varepsilon^{ij}\partial_j R\,. 
\label{ko1}
\eeq
As a result, $\frac{d\, R(x(\lambda))}{d \lambda}=0\,$. Thus, in a  generic spherically symmetric space-times, the areal radius $R$ is conserved along Kodama trajectories. In other words, a Kodama observer is characterized by the condition $R=R_0$. The operational interpretation goes as follows. Static observers in static black hole (BH) space-times become, in the dynamical case, Kodama observers whose velocity 
\beq
v^i_K=\frac{K^i}{\sqrt{\chi}}\,, \quad  \gamma_{ij}v^i_Kv^j_K=-1 \,.
\eeq 
The energy measured by this Kodama observer at fixed areal radius $R_0$ is
\beq
E=-v^i_K\partial_i I=-\frac{K^i \partial_i I}{\sqrt{\chi_0}}=\frac{\omega}{\sqrt{\chi_0}} \,,
\eeq
where $I$ is the classical action of the relativistic particle, $\omega=-K^i\partial_iI$ and $\partial_iI$ being its momentum. 
As a consequence, the tunneling rate  may be written also as
\beq
\Gamma  \simeq e^{-\frac{2 \pi}{ \kappa_H}\,\sqrt{\chi_0} E}\simeq e^{-\frac{E}{ T_0}}
\eeq
and the local quantity $T_0$
\beq
T_0=\frac{T_H}{\sqrt{\chi_0}}\,, \quad T_H=\frac{k_H}{2\pi}
\eeq
evaluated at radial radius $R_0$ is also invariant, since  it contains  the invariant factor $\sqrt{\chi}$.
In the static case $\chi=g^{rr}=-g_{00}$ and recalling   Tolman's theorem: 
``for a gravitational system at thermal equilibrium in a static gravitational field, the local temperature satisfies $T\sqrt{-g_{00}}=\mbox{constant} $''. As a consequence, $T_H=\frac{\kappa_H}{2\pi}$ is the intrinsic  temperature of the BH: the Hawking temperature. In the general static case, we confirmed this result by making use of the Unruh-de Witt detector formalism \cite{noi11}. In the dynamical case, the full interpretation is still missing, and one of the aims of this paper is to give a contribution in order to clarify this issue for cosmological horizons, by making use of concepts in linear quantum field theory.

Thus, in this paper, we shall evaluate the regularized vacuum expectation value, \textit{i.e.} the quantum vacuum fluctuation, given formally by  $\braket{\phi^2(x)}$, where $\phi(x)$ is a conformally coupled quantum field defined on a flat FLRW  space-time. This is an ill defined quantity and a regularization is necessary.  The computation of the coincidence limit will be done along world lines parametrized by proper time or arc length in the spacelike case. Therefore the final result will also  depend  on the invariant acceleration (the norm of the corresponding four-vector) of an arbitrary observer, and the role of Kodama observers will be investigated.  The computation of $\braket{\phi^2(x)}$ in a black hole space-time has also been used recently in \cite{Lowe:2013zxa} to discuss the reality of the firewall proposal around the black hole horizon and is conceivable that the analysis of these authors can be possibly relevant for the case of cosmic horizons as well.

Along the way, we shall also discuss the physical meaning of the renormalization procedures by insisting that they have to correspond to measurement procedures, and point out that many of them have no operational meaning in general.  We stress that the obtained results are  approximatively valid for all states for which the leading singularity is of Hadamard's type.\\
In some special cases, including the important static black holes, this renormalized fluctuation gives direct informations on the temperature associated with the quantum field at thermal equilibrium, and according to Buchholz \cite{buch} (see also the recent paper \cite{re}) in these cases one is dealing with a quantum thermometer. In general, the quantum fluctuation will still give informations on FLRW space-times. We then show that the fluctuations as measured locally by a co-moving observer are isotropic but they do not take the form of a quantum thermal bath with some characteristic horizon temperature parameter, as might be expected from general thermodynamical arguments based on horizon physics \cite{Padmanabhan:2009vy}.

The paper is organized as follows. In Section {\bf \ref{phi2}},  the vacuum fluctuation is introduced, including its formal renormalization. In Section {\bf \ref{appl}}, the general formula for the renormalized vacuum fluctuation is derived and a physical meaning is attached to the formal renormalization procedure. Some applications are also discussed in section {\bf \ref{kodama}}, where the peculiar class of Kodama observers is analyzed in relation to our problem. Conclusions are given in Section {\bf 5}.

\section{$<\phi^2>$ as an observable}
\label{phi2}
In the present section we discuss the role of the the quantum vacuum fluctuation, given formally by  $\langle\phi^2(x)\rangle$, where $\phi(x)$ is a quantum field defined on a generic curved space-time. It has been calculated in a cosmological context in Ref.~\cite{bunch,vilen,linde}.  A related quantity is the off-diagonal Wigthman function, given by
\begin{equation}
 W(x,x') = \langle \phi(x) \phi(x') \rangle
\end{equation}
and whose evaluation in the coincidence limit $x'\rightarrow x$ gives the fluctuation at space-time event $x$.  However, on general grounds, $W(x,x')$ in this limit possesses the so called Hadamard singularity: in $D=4$ one has (see for example \cite{brunetti})
\begin{equation}
 W(x,x') = \langle \phi(x) \phi(x') \rangle = \frac{1}{4 \pi^2} \frac{D(x,x')}{\sigma^2(x,x')}+V(x,x')\ln (\lambda\, \sigma^2(x,x'))+V(x,x')
\end{equation}
where $\sigma^2(x,x')$ is the geodesic distance between $x$ and $x'$, $\lambda$ is a characteristic dimensional parameter (a mass or a scalar curvature term) and $D, U$ and $V$ are smooth functions, regular at the coincidence limit.  We left understood the presence of the $i\, \epsilon$ factor which allows one to deal with tempered distributions. This means that in any case, at the coincidence limit, $W(x,x)$ is singular.

One of the simplest regularizations consists in removing the related Hadamard singularity associated with a reference space-time, typically Minkoswki space.  But in general the presence of the logarithmic divergence introduces a finite logarithmic ambiguity in the form of a dimensional parameter $\mu^2$. The structure of the Hadamard singularity depends on the geometry of the space-time and on the differential operator $L$ associated with the equation of motion of the field $\phi(x)$, while the finite part depends on the chosen quantum state. In our case, for sake of simplicity, we consider neutral quasi-free scalar fields, so that the operator $L$ consists in the D'Alembertian operator plus a  term which may depend on the gravitational coupling of the field and on its mass.  The renormalized value of the fluctuation $\langle\phi^2(x)\rangle_R$ contains physical information and in this sense it is an observable much simpler than the renormalized vacuum expectation value of the stress-energy tensor.

As an illustrative example, let us consider a free massive scalar field defined on the Euclidean manifold $S_1 \times R^3$, obtained by 
``rotating'' Minkowski time to imaginary values and then compactifying it with periodicity $\beta$.  It is well known that in this case one is dealing with a massive quantum scalar field in thermal equilibrium at temperature $T=\frac{1}{\beta}$.  The relevant operator  is
\begin{equation}
 L=-\partial^2_\tau-\nabla^2+M^2\,,
\end{equation}
$ \tau$ being the imaginary time with period $\beta$.

One may compute the regularized fluctuation by means of the zeta-function regularization (see for example \cite{h77,eli94,byt96} and references therein). The general formula is \cite{iellici,binzer}
\begin{equation}
\langle \phi(x)^2 \rangle_R = \lim_{\varepsilon \rightarrow 0}\left[\frac{d}{d \varepsilon}\Big(\varepsilon\, \zeta(1+\varepsilon,x)\Big) +   
\varepsilon\,  \zeta(1+\varepsilon,x)\, \ln \mu^2 \right]\,, 
\end{equation}  
where $\zeta(z,x)$ is the local zeta function associated with $L$ and $\mu^2$ is an arbitrary mass scale present when there is a pole of the local zeta function at $z=1$. We omit the details of the calculation, giving instead the result:
\begin{equation}
\langle \phi(x)^2 \rangle_R=\frac{M}{2\pi \beta}\sum_{n=1}^\infty \frac{K_1(n\beta M)}{n}+\frac{M^2}{8\pi^2}\ln\left( \frac{M^2}{\mu^2}\right) \,.
\end{equation}
Here $K_1(x)$ is the modified Bessel function. If $M$ is not vanishing, the thermal properties are not transparent.  Furthermore the ambiguity given by the arbitrary mass scale $\mu^2$ is still present and a physical renormalization precription is needed.  In the massless case, there is a drastic simplification and it is easy to show that the logarithmic term (with its arbitrary mass scale) is absent in the Hadamard singularity: the regularized result reads
\begin{equation}\label{T}
\langle \phi(x)^2 \rangle_R=\frac{1}{12 \beta^2}= \frac{T^2}{12}\,.
\end{equation}
Thus in this particular case the fluctuation acts as a quantum thermometer \cite{buch}. However, one should observe that in a generic curved space-time and for an arbitrary gravitational coupling a logarithmic term is always present in the Hadamard singularity, even though one is dealing with a massless scalar field: as a consequence of the regularization process a finite logarithmic term with an arbitrary mass scale $\mu^2$ is also present. However, if one restricts the analysis to the massless conformally coupled case, one may get rid of the logarithmic term (see for example \cite{candelas,ottewill}).

As we have just seen, the proposal put forward by Buchholz and collaborators seems to work for the massless scalar field at finite temperature on Minskoswki space-time. 
In \cite{buch} also the de Sitter space-time was investigated, and  we shall revisit this important case. For a Schwarzschild black hole the situation is not so simple, in the sense that the renormalized vacuum fluctuation still gives information on Hawking temperature but in a less direct way.  In fact, the result for the renormalized vacuum fluctuation of a massless conformally coupled scalar field on a Hartle-Hawking state reads \cite{candelas}

\begin{equation}
\langle \phi(x)^2 \rangle_R= \frac{T_H^2}{12V}-\frac{T_U^2}{12V}+T_H\Delta_H\,,
\end{equation}
where $T_H=\frac{1}{8\pi M}$ is the Hawking temperature, $T_U=\frac{M}{2\pi r^2 }$ is the Unruh temperature, $V=1-\frac{2M}{r}$ is the lapse function and $\Delta_H$ a finite contribution which can be numerically evaluated. It should be noted that on the horizon, since one is working on the 
Hartle-Hawking state, the vacuum fluctuation is finite, because the local (red-shifted) Hawking and Unruh temperature divergences cancel. Conversely, the finiteness of the correlation implies the value of the Hawking temperature.

Motivated by these arguments, we would like to consider a conformally coupled scalar field in a FLRW conformally flat space-time.  In this case the off-diagonal Wigthman function is given by (see, for example \cite{obadia,noi11})
\begin{equation}
 W(x,x') = \langle \phi(x) \phi(x') \rangle = \frac{1}{4 \pi^2} \frac{1}{\sigma^2(x,x')}\ ,
\end{equation}
where $\sigma^2(x,x') = a(t)a(t') \left( x-x' \right)^2$ and $a(t)$ is the scale factor.  In the coincidence limit $x \rightarrow x'$, one has the Hadamard singularity without the logarithmic term.

%A simple renormalization consists in subtracting the corresponding Minkowski contribution
%\begin{equation}\label{fp}
%  \langle \phi^2(x) \rangle_R =  \lim_{x' \rightarrow x}\ \frac{1}{4 \pi^2}\left( \frac{1}{\sigma^2(x,x')}- \frac{1}{\sigma^2_M(x,x')}\right)
%=F.P. \frac{1}{4 \pi^2}\left( \frac{1}{\sigma^2(x,x')}\right) \,.
%\end{equation}
%where $F.P.$ specifies that we are only considering the finite part resulting from the subtraction.

\section{A local expansion of $\braket{\phi^2(x)}$ in terms of parametrized world lines}
\label{appl}

As we argued in the previous section, massless conformally coupled fields could be a convenient probe to investigate at quantum level the properties of non trivial space-times through the quantity $\langle \phi^2(x) \rangle$.  In order to calculate this quantum fluctuation, one needs to evaluate to the inverse of the square geodesic distance between two events.  For our purposes, the geodesic distance may be conveniently expressed in terms of the corresponding world line $x(s)$ parametrized by the proper length, which is the proper time $\tau$ along timelike trajectories and the usual arc length along spacelike curves. This approach is similar to the one described in detail in the monograph \cite{BD} and it is the core of the adiabatic point-splitting regularization method. It is also strictly related to Unruh-De Witt detector approach (see for example \cite{BD} and references therein, and the references contained in the recent papers \cite{noi11,tanu13}). For a rigorous recent approach, see also \cite{Pina13}. In the following we shall discuss both timelike and spacelike correlations. We shall use both forms of the spatially flat FLRW metric
\beq
ds^{2}=-dt^{2}+a^{2}(t)(dr^{2}+r^{2}d\Omega^{2})=a^{2}(\eta)(-d\eta^{2}+
dr^{2}+r^{2}d\Omega^{2})
\eeq
where by a slight abuse of notation we denoted with the same symbol two really different functions, namely $a(t)$ and $a(\eta)=a(t(\eta))$; here $\eta=\int dt/a(t)$ is the conformal time.

\begin{center}\textit{Timelike correlations}\end{center}

The proper-time parametrized Wightman function reads
\begin{equation}
 W(x(\tau),x'(\tau')) = \frac{1}{4 \pi^2} \frac{1}{\sigma^2(\tau, \tau')}\ ,
\end{equation}
where
\begin{equation}\label{gener}
 \sigma^2(\tau,\tau') = a(\tau) a(\tau')\, \Big( x(\tau) - x(\tau') \Big)^2\,.
\end{equation}
is computed in the FLRW metric. Due to the isotropy of FLRW space-time we may restrict the analysis to radial trajectories, namely  $x(\tau)= \big(\eta(\tau),r(\tau)\big)$.  In order to discuss the coincidence limit, we put $\varepsilon=\tau-\tau'$.  An over dot will mean derivation w.r.t. proper time $\tau$. We define $a(\tau)=a(\eta(\tau))$ so that $\dot{a}=a^{2}H\dot{\eta}$, and so on.\\
It will be sufficient to make an expansion to fourth order in $\varepsilon$ of Eq.~(\ref{gener}), 
\begin{align}
\sigma^2(\tau,\tau-\varepsilon) \simeq & - \varepsilon^2\, a^2(\tau) \left( \dot{\eta}^{2} - \dot{r}^{2} \right)\nonumber
\\
& + \frac{1}{2} \varepsilon^3\, \partial_\tau \left[ a^2(\tau) \left( \dot{\eta}^{2} - \dot{r}^{2} \right) \right]\nonumber
\\
& + \frac{1}{12} \varepsilon^4\, \left[ 6 a\,\ddot{a}( \dot{r}^2  - \dot{\eta}^2) + 12 a\, \dot{a}(\dot{r}\, \ddot{r} -\, \dot{\eta}\, \ddot{\eta})\ + \right.\nonumber
\\
& \quad \quad \quad \quad \left.+\, 3\, a^2 (\ddot{r}^2 -  \ddot{\eta}^2) + 4\, a^2 (\dot{r}\, \dddot{r} - \dot{\eta}\, \dddot{\eta}) \right]\label{exp}
\end{align}
In order to simplify the expression, we just need to enforce the following relations:
\newline
i) $a^2 \dot{x}^2=-1$ (for timelike trajectories) and its derivatives w.r.t. $\tau$,\newline ii) the relation between cosmic time and conformal time $d\eta/dt = a^{-1}$ and \newline iii) $\dot{r}=a^{-1}\sqrt{\dot{t}^2-1}$,\newline

Because of (i) the coefficient of $\varepsilon^2$ is 1, while the coefficient of $\varepsilon^3$ is zero.  As regards the detailed calculations of the $\varepsilon^4$ term, see the Appendix \ref{app}.  The result in terms of $\dot{t}$ and $H(t)=\partial_ta(t) / a(t)$ is
\begin{equation}\nonumber
 \sigma^2(\tau, \varepsilon) = - \varepsilon^2 - \frac{1}{12} \left[ \frac{\ddot{t}^2}{\dot{t}^2-1} + 2\, \ddot{t}\, H + \dot{t}^2 \left( H^2 + 2\, \partial_t H \right) \right] \varepsilon^4 + O( \varepsilon^6)
\end{equation}
Note that $\dot{t}$ fully determines the trajectory via (iii), while $H(t)$ is determined by $a(t)$ (\textit{i.e.} by the model one is dealing with).  Furthermore, the $\tau$-dependent coefficient in square brackets can be rewritten in a more enlightening way:
\begin{equation}\label{expa}
 \sigma^2(\tau, \varepsilon ) = - \varepsilon^2 - \frac{1}{12} \Big[ A^2 + H^2 + 2 \dot{t}^2 \partial_t H \Big] \varepsilon^4 + O(\varepsilon^6)
\end{equation}
where
\begin{equation}\nonumber
 A^2 = \left[ \frac{\ddot{t}}{\sqrt{\dot{t}^2-1}} +  H \sqrt{ \dot{t}^2-1}  \right]^2
\end{equation}
is the square of the four-acceleration along the trajectory.

At this point we must discuss a crucial matter: the renormalization of the singular term in the Wightman function. We think that, in order to avoid any possible ambiguity, this should be done by making reference to the actual method of measurement. It seems there are many possibilities, but to keep matter as plain as possible we will consider two simple and sensible ways to define the finite part of the correlation. The first is to subtract the value for Minkowski space on a linear trajectory, a flat geodesics in fact; the second is to subtract the value on the actual trajectory as embedded in flat space. Thus in the first case we set the detector so that it registers no signal when it is at rest in a freely falling frame; in the second case we set it so that it gives no signal when it is moving on the actual trajectory again in a freely falling frame. If the detector is small with respect to the curvature scale, for example a point-like monopole detector, well known arguments based on the equivalence principle will imply that the detector in general will register a signal when it is moving or is at rest in a comoving frame. Since Minkowski space can be locally reached by passing to a freely falling frame, both prescriptions are in principle achievable. By contrast, any subtraction corresponding to a globally non isometric space-time has no operational meaning in the given space-time, and should not be used. For instance, we can give $\braket{\phi^{2}}$ any value we like by subtracting its un-renormalized value in a contracting universe, but clearly there is no natural physical meaning in this entirely arbitrary procedure.\\
In the case of inertial trajectories in Minkowski space-time one has $\dot{x}^{\mu}=U^{\mu}\tau$, $U^{2}=-1$ and $a(t)=1$. Thus,
\begin{equation}\label{expam}
 \sigma_M^2(\tau, \varepsilon ) = - \varepsilon^2\,. 
\end{equation}
The renormalization following from the first prescription  requires the subtraction of this contribution from the expression given by Eq.~(\ref{expa}).  The renormalized vacuum fluctuation is then given by 
\begin{equation}\label{temper0}
\langle \phi^2(x) \rangle_R  = \frac{1}{48 \pi^2} \left[ \frac{\ddot{t}^2}{\dot{t}^2-1} + 2\, \ddot{t}\, H + \dot{t}^2 \left( H^2 + 2\, \partial_t H \right) \right]\,, 
\end{equation}
or the equivalent form 
\begin{equation}\label{temper}
\langle \phi^2(x) \rangle_R  = \frac{1}{48 \pi^2} \left[ A^2 + H^2 + 2 \dot{t}^2 \partial_t H \right]
\end{equation}
This result is in agreement with  the one obtained by Obadia \cite{obadia} within the Unruh-deWitt detector approach.
Furthermore, in this form, it shows in a clear way the contribution coming from the proper motion along the trajectory (through $A$ and hence $\dot{t}$) and the one coming from the dynamics of the cosmological model (through $H$).  The third term in square brackets represents a mixed contribution, which vanishes for stationary cosmological models.\\
Thus adjusting the detector so that it does not register for inertial motion in flat space provides a probe of certain features of space-time geometry and of the actual motion. Most interesting of course would be the proper motion of our neighborhood relative to the Hubble flow.

Interesting is the case whereby $\dot{H}\ll H^{2}$: then $\braket{\phi^{2}}$ has a thermal interpretation in terms of de Sitter temperature (even in the presence of some acceleration, see below). In the general case there is not such an interpretation. For instance in the Einstein-\ds{} model $\dot{H}=-3H^{2}/2$ so that $\braket{\phi^{2}}_R=-H^{2}/24\pi^{2}$ despite the presence of a trapping horizon.\\
For a co-moving observer (we will see that this is a special class of Kodama observers)  Eq.~\eqref{temper} reduces to
\begin{equation}\label{tempercomov}
\langle \phi^2(x) \rangle_R  = \frac{1}{48 \pi^2} \left[H^2 + 2\dot{H} \right]
\end{equation}
Evidently the correlation acts as a probe for testing both the Hubble parameter as well as its time derivative. In de Sitter space relative to inflationary coordinates, one has  $\dot{H}=0$ and the correlation measures the Gibbons-Hawking temperature of de Sitter space \cite{GH} for a geodesic observer (see Eq.~\eqref{T}). The renormalization prescription just adopted should be equivalent to a subtraction relative to the Bunch-Davies vacuum, the same to be used in inflation theory, and in this sense is the favorite one, though the conformal field is not very relevant in inflationary theory.  But there are other possibilities: an appropriate one for massless fields is perhaps the Kirsten-Garriga vacuum \cite{Kirsten:1993ug}, or the various \ds{} $\alpha$-vacua. We have not investigated this matter any more.

The second prescription consists in  subtracting the fluctuation on the actual trajectory as embedded in flat Minkowski space, namely  Eq.~\eqref{temper0} with $H=0$. As a result, 
\begin{equation}\label{temper01}
\langle \phi^2(x) \rangle_R  = \frac{1}{48 \pi^2} \left[ 2\, \ddot{t}\, H + \dot{t}^2 \left( H^2 + 2\, \partial_t H \right) \right]\,, 
\end{equation}
 It is designed to separate  the Unruh effect from the expansion, but a coupling $\ddot{t}H$ actually remains. It resembles a Machian effect, showing that subtracting the acceleration relative to absolute space, as would be expected in this case, leaves nonetheless a coupling with the whole universe.  It is clearly not possible to cancel this term by renormalization with respect to a freely falling frame, because in such a frame the rest of the universe disappears from the stage.

In spherical symmetry the physical radial trajectory  is identified by the variation of the areal radius $R(t) = r\, a(t)$ so that, making use of the line element Eq.~(\ref{metric}) restricted on a radial path,
\begin{equation}\label{tdot}
 \dot{t} = \frac{1}{\sqrt{1-\left( \frac{dR}{dt} - \frac{R}{R_H} \right)^2}}
\end{equation}
and defining $\mathcal{V}= \frac{dR}{dt}- \frac{R}{R_H}$, one can express Eq.~(\ref{temper}) as
\begin{equation}\label{finalphi}
\langle \phi^2(x) \rangle_R =  \frac{1}{48 \pi^2}\left[   H^2 + \frac{1}{1-\mathcal{V}^2} \left( \frac{\dot{\mathcal{V}}}{\sqrt{1-\mathcal{V}^2}} + H\, \mathcal{V}\right)^2 +  2 \frac{\partial_t H}{1-\mathcal{V}^2} \right]
\end{equation}
where the second term is the proper acceleration.  This expression can also be rewritten in an alternative form
\begin{equation}
\langle \phi^2(x) \rangle_R =  \frac{1}{48 \pi^2}\, \frac{1}{1-\mathcal{V}^2}\left[ H^2 + 2 \partial_t H + \frac{1}{\sqrt{1-\mathcal{V}^2}} 
\left( \frac{\dot{\mathcal{V}}^2}{\sqrt{1-\mathcal{V}^2}} + 2 H\, \mathcal{V}\dot{\mathcal{V}} \right)\right]\,.
\label{g1}
\end{equation}
in view of the discussion  in the following sections.

\begin{center}\textit{Spacelike correlations}\end{center}

In this case we use as parameter the arc length $s$ along the curve and we will simplify matters by choosing a purely spatial trajectory within the cosmic time slices of the metric.  
A  simple calculation gives now nothing interesting, namely the exact result
\beq\label{correl}
\sigma^{2}(s,\epsilon)=\epsilon^{2}
\eeq
Evidently we need to consider spacelike trajectories with some extension in time. Formally we can get the spacelike result by taking imaginary proper time, say $\tau=is$, in Eq.~\eqref{temper0}; we obtain, isolating the acceleration from the expansion
\beq\label{correlT}
\sigma^{2}(s,\epsilon)=\frac{1}{48 \pi^2} \left[ \frac{t^{''2}}{t^{'2}+1} - 2\, t^{''}\, H - t^{'2} \left( H^2 + 2\, \partial_t H \right) \right]+\cdots
\eeq
where a prime denotes $d/ds$. Adopting once more the renormalization with respect to Minkowski space, we see that the spacelike correlation are again a probe of the universe expansion. But unlike the timelike case, this time there is no detector available and no obvious local method to measure them\footnote{In cosmology the power spectrum is first determined by observations in causal directions, then the spatial correlations are reconstructed via Fourier transform.}.

\section{Kodama trajectories}
\label{kodama}

The previous results did not specify any special trajectory. Here we restrict our analysis to the class of Kodama observers, {\it i.e.} the observers characterized by the condition $a(t)r\equiv R(t)=R_0$.  The importance of these observers is related to the properties of the Kodama vector field  in spherical symmetry (see the Introduction). In the static patch of \ds{} space or static spaces they stay at constant $r$, in FLRW spaces they are on top of constant areal radius, namely $R=r(t)a(t)$ is a constant, and become null on the Hubble sphere, in Rindler space they correspond to uniformly accelerated observers in Minkowski space, and so on. We are interested to Kodama trajectories since they are the closest analogue of the  hyperbolic histories giving rise to characteristic thermal effects. \\
From Eq.~(\ref{finalphi}) we find
\beq\label{T2}
\langle \phi^2(x) \rangle_R = \frac{1}{48 \pi^2}\frac{1}{1-R_0^2 H^2}\left(H^2+2 \partial_{t}H+\frac{(\partial_{t}H)^{2}R_{0}^{2}}{\left(1-R_0^2 H^2\right)^{2}} + \frac{2H^{2}R_{0}^{2}\partial_{t}H}{\sqrt{1-H^{2}R_{0}^{2}}}\right)\,.
\eeq
which for the comoving observer at $R_{0}=0$ reduces to Eq.~\eqref{tempercomov}.
Apparently we see no natural thermal interpretation of this formula. \\
An expression of the characteristic surface gravity associated with a trapping horizon in cosmology is given by Eq.~\eqref{H}, which when redshifted to a Kodama trajectory takes the form
\beq
\kappa_{H}=\left(H+\frac{\dot{H}}{2H}\right)(1-H^{2}R_{0}^{2})^{-1/2}
\eeq
According to a possible interpretation, this time-dependent parameter controls the particle creation rate by the trapping horizon located at $r_{H}a=H^{-1}$, as given by Eq.~\eqref{g}, lending support to the interpretation of $\kappa_{H}/2\pi$ as an effective temperature parameter. 
But \eqref{T2}, though similar, is quite different from  \eqref{T} at the base of Buchholz's proposal for this temperature. Evidently the cosmological horizon does not create a thermal background radiation with a simple geometric description, as it happens to black holes, unless $\dot{H}=0$ (this case will be considered in the next subsection).

It seems truly remarkable that a local measurement such as the one involved in the fluctuation has a geometrical interpretation in terms of a quantity which is associated to a very distant, observer dependent Hubble sphere, and it seems at odds with general results concerning the effects of expansion on local systems. Is this another instance of a Machian effect in cosmology? More modestly, we could simply say that there is at least a renormalization prescription which produces a geometrically meaningful (i.e. tightly connected to space-time geometry) result. But we have to recall that the expectation value is not a truly local quantity because the vacuum state to which it refers actually probes a cosmic time slice in its entirety.

\subsection{Stationary space-times}

These observations bring us, as an important check, to consider de Sitter space, which has $H(t)=H_0$ constant in inflationary coordinates\footnote{These are the only ones with flat spatial sections to which our formalism apply.}.  The expression simplifies to
\begin{equation}\label{ds}
\langle \phi^2(x) \rangle_R = \frac{1}{48 \pi^2}\frac{H_0^2}{1-R_0^2 H_0^2}\,.
\end{equation}
thus  one has
\begin{equation}\label{ds0}
 \langle \phi^2(x) \rangle_R = \frac{1}{12}\frac{T_{GH}^2}{(1-R_0^2 H_0^2)}\,, 
\end{equation}
 where $T_{GH}=\frac{H_0}{2\pi}$ is the Gibbons-Hawking temperature \cite{GH}, $(1-R_0^2 H_0^2)$ being the kinematical red-shift  factor. 
One also has the known result \cite{thirring}
\begin{equation}\label{ds1}
 \langle \phi^2(x) \rangle_R = \frac{1}{12}\left(\frac{A_0^2}{4\pi^2}+T^2_{GH}\right)\,,
\end{equation}
where 
\[
A_0=\frac{R_{0}H^{2}}{\sqrt{1-H^{2}R_{0}^{2}}}
\]
is the invariant acceleration of the Kodama observer (needed to remain at constant radial distance from the origin in the static frame).

One may obtain a confirmation of the above result by observing that the de Sitter space-time admits a static patch. With regard to this issue, we may discuss the black hole general case .  In fact, a generic spherically symmetric static black hole metric reads
\beq
ds^2
 &=&-V(r)dt_S^2+\frac{dr^2}{V(r)}+r^2 d \Omega^2 \nonumber \\
&=&V(r^*)[-dt_S^2+(dr^{*})^2]+r^2(r^*) d \Omega^2 \;.
\label{s*} 
\eeq 
where $t_S$ is the time coordinate in the static patch and $r^*$ is the tortoise coordinate given by $dr^*(r)=\frac{d r}{V(r)}$. Introducing the Kruskal-like coordinates defined by
\beq
X=\frac{1}{\kappa_H}e^{\kappa r^*}\cosh(\kappa_H t_S)\,, \quad T=\frac{1}{\kappa_H}e^{\kappa_H r^*}\sinh(\kappa_H t_S)\,,
\eeq
where $\kappa_H=\frac{V'_H}{2}$ is the usual Killing surface gravity, one obtains
\beq
ds^2 = e^{-2\kappa_H r^*}\,V(r^*)[-dT^2+dX^2]+r^2(T,R) d \Omega^2\,, 
\label{ks*} 
\eeq
where now $r^*=r^*(T,X)$.  The key point to recall here is that in the Kruskal gauge the normal metric  is conformally related to two-dimensional Minkoswki space-time. The second observation is that Kodama observers are defined by the integral curves associated with the Kodama vector, thus the areal radius $r$ and $r^*$ are {\it constant}, say $r=r_0$. As a consequence, one is dealing with an effective flat FLRW space-time
\beq
ds^2=V_0e^{-2\kappa_H r_0^*}(-dT^2+dX^2)=-dt^2+a^2(r_0)dX^2\,,
\label{BHf}
\eeq
where $t=\sqrt{V_0}e^{-\kappa_H r_0^*}T$ is a new ``cosmological'' time, and $a(r_0^*)=\sqrt{V_0}e^{-\kappa_H r_0^*}$ is the related constant scale factor.
Furthermore, the proper time along Kodama trajectories reads
\beq
d\tau^2=V_0 dt_S^2=V_0e^{-2\kappa_H r_0^*}(dT^2-dX^2)=dt^2-a^2(r_0)dX^2\,,
\eeq
Finally one also has, as functions of the proper time,
\beq
X(\tau)&=&\frac{1}{\kappa_H}e^{\kappa_H r_0^*}\cosh\left(\kappa_H\frac{\tau}{\sqrt{V_0}}\right) \nonumber\\
T(\tau)&=&\frac{1}{\kappa_H}e^{\kappa_H r_0^*}\sinh \left(\kappa_H \frac{\tau}{\sqrt{V_0}} \right)\,.\label{ bestiale}
\eeq 
Since for the metric Eq.~(\ref{BHf}) one has $H\equiv \partial_t a / a=0$, we have to apply the general formula (\ref{temper0}), namely the one associated with the first renormalization prescription,  obtaining
\begin{equation}\label{temper0bh}
\langle \phi^2(x) \rangle_R  = \frac{1}{48 \pi^2} \left[ \frac{\ddot{t}^2}{\dot{t}^2-1} \right]\,. 
\end{equation}
For the static black hole one obtains
 \begin{equation}\label{tddot}
\dot{t}  =  \cosh\left(\kappa_H\frac{\tau}{\sqrt{V_0}}\right)\,, \quad 
\ddot{t}= \frac{\kappa_H}{\sqrt{V_0}}\sinh\left(\kappa_H\frac{\tau}{\sqrt{V_0}}\right)\,.
\end{equation}
which gives for the  quantum fluctuation
\begin{equation}\label{temper0bhf}
\langle \phi^2(x) \rangle_R  = \frac{1}{48 \pi^2}  \frac{\kappa_H^2}{V_0}=\frac{T_H^2}{12\ V_0}\,. 
\end{equation}
As a result, also in this case, one has a Buchholz quantum thermometer with Hawking temperature at infinity $T_H=\frac{\kappa_H}{2\pi}$ redshifted by the usual Tolman factor $V_0$.
For de Sitter $V(r)=1-H_0^2r^2$ and $T_H=\frac{|\kappa_H|}{2\pi}=\frac{H_0}{2\pi }$, thus recovering the previous result. We may conclude that this temperature is an intrinsic property of \ds{} space, not depending on the coordinates used\footnote{From a canonical perspective it is an observable in the Bergmann sense of  a ``gauge invariant phase space function''.}.

\subsection{Non stationary space-times: the Big Rip}

For the Big Rip solution \cite{cald,br},  one has
\beq
H(t)=\frac{c}{(t_s-t)^{\alpha}}\,, \quad c>0\ \text{and}\ \alpha>0\, ,
\eeq
As in stationary scenarios, for Kodama observers the fluctuation diverges on the Hubble horizon $R_H = H^{-1}$.  On the other hand, if $0<R_0<R_H$, with every $\alpha > 1/2$ the fluctuation at the Big Rip attains a finite value:
\begin{equation}
 \lim_{t \rightarrow t_s} \langle \phi^2(x) \rangle_R   =- \frac{1}{48 \pi^2} \frac{1}{R_0^2}\label{T2br2}
\end{equation}
If $\alpha=1/2$ we have instead
\begin{equation}
 \lim_{t \rightarrow t_s} \langle \phi^2(x) \rangle_R   = -\frac{1}{192\pi^2\, c^4\, R_0^4} - \frac{1}{48 \pi^2} \frac{1}{R_0^2}\label{T2br2a}
\end{equation}
It is interesting to note that only in this latter case the limiting value depends on $(c\, R_0)^{-4}$, \textit{i.e.} on the details of the model through $c$.

The finite, but negative values of $\langle \phi^2(x) \rangle_R$ at the Big Rip is due to the fact that a Kodama observer with $R_0\neq0$ has a non-vanishing proper acceleration which diverges towards the singularity: this divergence offsets exactly the infinite contribution coming from $H(t)$.

\section{Conclusions}

We have computed at length the simplest local observable available in a conformal scalar field theory in a flat FLRW metric. Instead of throwing away the full short distance singularity of its Wightman function, as one might have supposed to do, we retain some terms which look meaningful. We have given justifications for this procedure. In \ds{} space it gives the result obtained in inflation theory for the field in the Bunch-Davies vacuum. The fluctuations as computed shows no relation with the thermodynamics of cosmic horizons, except for \ds{} space or in a quasi-de Sitter regime. The Machian flavor of the results, namely the connection of local measurements with the expansion of the universe, is explained by the non local character of the vacuum state. 

\appendix
\section{Appendix: fourth-order term in the expansion of $\sigma^2$}
\label{app}

Let us consider the $\varepsilon^4$ term in Eq.~(\ref{exp}):
\begin{align}
 &\frac{1}{12} \left[ 6\, a\, \ddot{a} \left( -\dot{\eta}^2 + \dot{r}^2 \right) + 12\, a\, \dot{a} \left( -\dot{\eta}\, \ddot{\eta} + \dot{r}\, \ddot{r} \right) \right. + \nonumber
 \\
 & \quad \quad \left. +\, 3\, a^2 \left( -\ddot{\eta}^2 + \ddot{r}^2 \right) +4\, a^2 \left( -\dot{\eta}\, \dddot{\eta} + \dot{r}\, \dddot{r} \right) \right]\nonumber
\end{align}
\begin{enumerate}
 \item the first term can be re-written by using the condition $a^2 \dot{x}^2=-1$.  The result is
 \begin{equation*}
  6\, a\, \ddot{a} \left( -\dot{\eta}^2 + \dot{r}^2 \right)\ =\ -6\, \frac{\ddot{a}}{a}\ =\ -6\, \dot{t} \left( \ddot{t}\, \partial_t{H} + \dot{t} \right)
 \end{equation*}
 \item by using the fact that $\partial_{\tau}\left[ a^2 \dot{x}^2 \right] = 0$, the second term becomes
 \begin{equation*}
  12\, a\, \dot{a} \left( -\dot{\eta}\, \ddot{\eta} + \dot{r}\, \ddot{r} \right) = 12 H^2\, \dot{t}^2
 \end{equation*}
 \item by deriving once more the condition $\partial^2_{\tau}\left[ a^2 \dot{x}^2 \right] = 0$, the sum of the last two terms can be recast in the form
 \begin{align*}
  &3\, a^2 \left( -\ddot{\eta}^2 + \ddot{r}^2 \right) +4\, a^2 \left( -\dot{\eta}\, \dddot{\eta} + \dot{r}\, \dddot{r} \right) =
  \\
  = &-H^2\, \dot{t}^2 + 4 H\, \ddot{t} - \frac{\ddot{t}}{\dot{t}^2-1} + 4\dot{t}^2\, \partial_tH
 \end{align*}
\end{enumerate}
Putting all together, the $\varepsilon^4$ term is given by
\begin{equation}
 -\frac{1}{12} \left[ \frac{\ddot{t}^2}{\dot{t}^2-1} + 2\, H\, \ddot{t} + \dot{t}^2 \left( H^2 + 2\, \partial_tH \right) \right]\nonumber
\end{equation}

\end{document}